\documentstyle[twocolumn,floats,aps,epsf]{revtex}
\begin{document}
\twocolumn[
\hsize\textwidth\columnwidth\hsize\csname@twocolumnfalse%
\endcsname
\draft
\title{Phase Diagrams of Mixed Spin Chains with Period 4 \\ 
by the Nonlinear $\sigma$ Model}
\author{Ken'ichi Takano \\
Laboratory of Theoretical Condensed Matter Physics and \\
Research Center for Advanced Photon Technology, \\
Toyota Technological Institute, Nagoya 468-8511, Japan}
%
\date{Received}
\maketitle
\begin{abstract}
      We study mixed quantum spin chains consisting of two kinds 
of spins with magnitudes, $s_a$ and $s_b$. 
      The spins are arrayed as $s_a$-$s_a$-$s_b$-$s_b$ in a unit cell 
and the exchange couplings are accordingly periodic with period 4. 
      The spin Hamiltonian is mapped onto a nonlinear $\sigma$ 
model based on the general formula for periodic inhomogeneous 
spin chains. 
      The gapless condition given by the nonlinear $\sigma$ model 
determines boundaries between disordered  phases in the space of 
the exchange parameters. 
      The phase diagram has a rich phase structure characterized by 
the values of $s_a$ and $s_b$. 
      We explain all phases in the singlet-cluster-solid picture 
which is an extension of the valence-bond-solid picture. 
\end{abstract}
]


\narrowtext

\section{Introduction}

      Quantum spin systems have been investigated for long time 
and have revealed quantum states which do not appear in the 
corresponding classical systems. 
      They have rich content as quantum many-body systems, 
although they are described by Hamiltonians much simpler than 
itinerant electron systems. 
      Among spin systems the spin chain is especially interesting 
because quantum fluctuation is the largest in one dimension. 
      For the last two decades, one of the most exciting discoveries 
in the spin chain is the Haldane gap. 
      That is, a homogeneous antiferromagnetic spin chain is gapful 
if the spin magnitude is an integer, and is gapless if it is 
a half-odd integer. 
      Haldane proposed this prediction based on the mapping of a spin 
chain onto the nonlinear $\sigma$ model (NLSM) in a semiclassical 
manner \cite{Haldane}. 
      Since then, the NLSM is strongly recognized to be important 
in research of quantum spin systems. 
      Various interesting aspects of the NLSM method for spin 
chains are found in Refs. \cite{Affleck,Fradkin,Tsvelik}. 

      Affleck reformulated the NLSM method in an operator 
formalism \cite{Affleck2}. 
      He divided a spin chain into spin pairs and transformed the 
spin operators for each spin pair into operators representing 
an antiferromagnetic motion and a small fluctuation. 
      The NLSM is analyzed by the field theoretic method 
\cite{Affleck3}. 
      It is important that his operator formalism is applicable 
even to a spin chain with bond alternation. 
      The spin chain with spin magnitude $\frac{1}{2}$ has 
a finite spin gap if the bond alternation is finite. 
      In this case, the ground state is in a dimer phase. 
      A gapless spin excitation can appear only at the phase boundary 
between dimerized phases, where the system has no bond alternation. 
      In contrast, a spin chain with spin magnitude 1 has a gapless 
spin excitation at a finite strength of the bond alternation. 
      These results consists with those of numerical 
calculations \cite{Kato,Yamamoto} and with experimental 
results \cite{Hagiwara}. 

      The bond alternation is the simplest inhomogeneity for 
the exchange parameter; the spin chain with bond alternation 
is a system with period 2. 
      The next interest is in inhomogeneous spin chains with 
periods more than 2. 
      In general we can consider two kinds of inhomogeneities 
in a spin chain: 
      one is for the exchange parameter and the other is for the 
spin magnitude. 
      It is difficult to extend the operator formalism of Affleck 
to various inhomogeneous cases, since a very complicated 
transformation among operators may be needed. 
      In particular, the difficulty is serious for mixed spin chains, 
where the spin magnitude is inhomogeneous. 

      In a previous paper, we unambiguously derived the NLSM for 
the general mixed spin chain with finite period \cite{Takano1}. 
      The derivation is based on dividing the spin chain into blocks 
in a path integral formalism. 
      Each block is a single unit cell or consists of a series of unit 
cells. 
      A spin pair used by Affleck \cite{Affleck2} can be regarded 
as an especially simple block containing only two spins. 
      We carried out a transformation for spin variables in each 
block so as to preserve the original degrees of freedom. 
      We finally obtained the NLSM action describing low energy 
behavior of the original spin chain \cite{Fukui1}. 
      The topological angle in the topological term of the NLSM 
depends on the exchange parameters and the magnitudes of 
spins, and determines whether or not the system has a gapless 
spin excitation. 
      The gapless condition is given in a closed equation form; 
i.~e. the {\it gapless equation}. 
      This gapless equation is quite general, and is applicable 
to various spin chains. 

      In this paper, we present and examine the phase diagrams 
of spin chains with period 4 by means of the NLSM. 
      We restrict ourselves to the case that two kinds of spins, 
$s_a$ and $s_b$, compose a spin chain. 
      The condition that the ground state is singlet allows only the 
array of $s_a$-$s_a$-$s_b$-$s_b$ in a unit cell. 
      We apply the general NLSM method to the spin chain with 
period 4 and write down the gapless equation. 
      The gapless equation determines gapless phase boundaries 
and hence the phase diagram for each pair of $s_a$ and $s_b$. 
      The phase diagram generally consists of many phases 
when $s_a$ and/or $s_b$ are large. 
      To understand these phases, we introduce the 
singlet-cluster-solid (SCS) picture. 
      It is an extension of the valence-bond-solid (VBS) picture 
\cite{Affleck4} and its special versions have been used 
(e.~g. Refs. \cite{Takano2,Chen1}). 
      The SCS picture systematically explains all phases for any 
values of $s_a$ and $s_b$. 

      This paper is organized as follows. 
      In Sec.~II, we introduce the general Hamiltonian for spin 
chains with period 4 and with singlet ground state. 
      In Sec.~III, the NLSM for the general spin chain with any period 
is reviewed. 
      The gapless equation determined by the topological term is 
written. 
      In Sec.~IV, the general gapless equation is specialized to the 
period-4 case and its properties are mentioned. 
      In Sec.~V, the case of homogeneous spin magnitudes 
($s_a$=$s_b$) is examined. 
      Phase diagrams are obtained and explained in the SCS picture. 
      In Sec.~VI, the phase diagram is examined in the general case 
($s_a$$<$$s_b$) and is interpreted in the SCS picture. 
      Section VII is devoted to summary and discussion.

\section{Mixed Spin Chain with Period 4} 

     The Hamiltonian with period 4 is generally written as 
\begin{eqnarray}
\label{Hamiltonian}
      H = \sum_{j=1}^{N/4} &{}& 
 ( J_1 \, {\bf S}_{4j+1} \cdot {\bf S}_{4j+2} 
 + J_2 \, {\bf S}_{4j+2} \cdot {\bf S}_{4j+3} 
\nonumber \\
 &{}&+ J_3 \, {\bf S}_{4j+3} \cdot {\bf S}_{4j+4} 
 + J_4 \, {\bf S}_{4j+4} \cdot {\bf S}_{4j+5} ) , 
\end{eqnarray}
where ${\bf S}_j$ is the spin at site $j$ with magnitude $s_j$. 
      The number of lattice sites is $N$, the lattice spacing 
is $a$ and the system size is $L = a N$. 
      We only consider the antiferromagnetic exchange interaction 
($J_i > 0$). 
      Since the system is periodic with period 4 in the spin magnitude 
as well as in the exchange parameter, we generally have 4 different
spin magnitudes: $s_1$, $s_2$, $s_3$ and $s_4$ in a unit cell. 
      However we concentrate systems with singlet ground states. 
      Following the Lieb-Mattis theorem \cite{Lieb}, a singlet ground 
state is realized if the system satisfies the condition 
\begin{equation}
\label{restriction}
       s_1 - s_2 + s_3 - s_4 = 0 ; 
\end{equation}
otherwise the system has a ferrimagnetic ground state. 

\begin{figure}[btp]
\begin{center}\leavevmode
\epsfxsize=60mm
\epsfbox{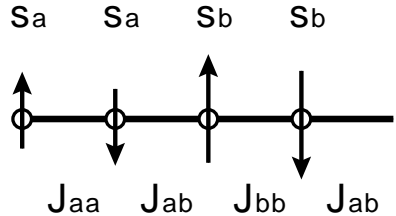}
\caption{Spin magnitudes and exchange parameters in a unit cell 
of the Hamiltonian (\ref{Hamiltonian}) with   Eqs. (\ref{magnitude}) 
and (\ref{exchange}).}
\label{unit_cell}
\end{center}
\end{figure}

       We treat the case that two kinds of spins are mixed and 
denote the magnitudes as $s_a$ and $s_b$. 
       To consist with the restriction (\ref{restriction}), the spin
magnitudes in the $j$th unit cell are taken as 
\begin{eqnarray}
\label{magnitude}
s_a &\equiv& s_{4j+1} = s_{4j+2} , \nonumber \\ 
s_b &\equiv& s_{4j+3} = s_{4j+4} \quad (s_a \le s_b)
\end{eqnarray}
without further loss of generality. 
       Correspondingly, we take the exchange parameters as 
\begin{equation}
\label{exchange}
J_{aa} \equiv J_1 , \quad J_{ab} \equiv J_2 = J_4 , \quad 
J_{bb} \equiv J_3 . 
\end{equation}
      We often use the unit of $J_{bb} (= J_3) = 1$. 
      A unit cell of the lattice is illustrated in Fig.~\ref{unit_cell}. 

\section{NLSM for the general spin chain}

      In this section we summarize the essential results in Ref. 
\cite{Takano1},  in which the NLSM is derived 
for the general spin chain with arbitrary period. 
      The Hamiltonian examined in Ref. \cite{Takano1} is 
\begin{equation}
      H = \sum_{j=1}^{N} J_j \, {\bf S}_j \cdot {\bf S}_{j+1} 
\label{general_Hamiltonian}
\end{equation}
with 
\begin{equation}
      J_{j+2b} = J_j , \quad s_{j+2b} = s_j , 
\label{period}
\end{equation}
where $b$ is a positive integer. 
      We consider the spin chain by dividing it into blocks 
of size $2b$. 
      The period $2b$ need not be a unit cell, but 
it may be a positive integer multiple of a unit cell 
\cite{even_block}. 
      In order that the mapping of the Hamiltonian 
(\ref{general_Hamiltonian}) to the NLSM is successfully carried 
out, the spin magnitudes must satisfy the restriction 
\begin{equation}
      \sum_{j=1}^{2b} (-1)^j s_{j} = 0 . 
\label{restriction_s}
\end{equation}
      The restriction coincides with the condition (\ref{restriction}), 
which excludes the ferrimagnetic ground state. 

      The expectation value of a spin operator in a coherent state is
expressed as 
\begin{equation}
           <{\bf S}_j> = (-1)^j s_j {\bf n}_j 
\label{coherent}
\end{equation}
with a unit vector ${\bf n}_j$. 
      The partition function $Z$ at temperature $1/\beta$ is 
then represented as 
\begin{eqnarray}
      Z &=& \int D[{\bf n}_j] \, \prod_j \delta({\bf n}^2_j - 1) \, e^{-S} , 
\label{partition}
\\
      S &=& i \sum_{j=1}^{N} (-1)^j s_j w[{\bf n}_j] 
\nonumber \\ 
         &+& \int_0^{\beta} d\tau \sum_{j=1}^{N} 
                J_j s_j s_{j+1} {\bf n}_j \cdot {\bf n}_{j+1} . 
\label{action_S} 
\end{eqnarray}
      In the action (\ref{action_S}), the first term comes from the 
Berry phase and 
$w[{\bf n}_j]$ is the solid angle which the unit  vector ${\bf n}_j$
forms in the period $\beta$. 

      We transform the spin variables \{${\bf n}_{j}$\} into gradually 
changing unit vectors \{${\bf m}(p)$\} and their small fluctuations 
\{${\bf L}_q(p)$\}. 
      The spin variable at the $q$th site in the $p$th block is written as 
\begin{eqnarray}
      {\bf n}_{2bp+q} = (1 - z_q) {\bf m}(p) + z_q {\bf m}(p-\gamma_q) 
                                 + a {\bf L}_q(p)  
\label{transform}
\end{eqnarray}
with $z_q = |b-q|/2b$ and $\gamma_q = {\rm sgn}(b-q)$. 
       This transformation does not change the number of the original 
degrees of freedom as is explained in Ref. \cite{Takano1}. 
      Integrating out the fluctuations and taking the continuum limit, 
we finally obtain the effective action \cite{Takano1}: 
\begin{eqnarray}
      S_{\rm eff} = \int^{\beta}_0 d\tau \int^{L}_0 dx 
\biggl\{ 
      &-& i \frac{J^{(0)}}{J^{(1)}} 
{\bf m} \cdot (\partial_{\tau} {\bf m} \times \partial_x {\bf m}) 
\nonumber \\ 
      &+& \frac{1}{2aJ^{(1)}} \biggl( 
\frac{J^{(1)}}{J^{(2)}} - \frac{J^{(0)}}{J^{(1)}} \biggl) 
(\partial_{\tau} {\bf m})^2 
\nonumber \\ 
      &+& \frac{a}{2} J^{(0)} (\partial_x {\bf m})^2 
\biggl\} , 
\label{action-NLSM}
\end{eqnarray}
where $J^{(n)}$ ($n$ = 0, 1, 2) is defined as 
\begin{eqnarray}
\label{def_J_n}
\frac{1}{J^{(n)}} &=& \frac{1}{2b} \sum_{q=1}^{2b} 
\frac{({\tilde s}_q)^n}{J_q s_q s_{q+1}} , \\
      {\tilde s}_q &=& \sum_{k=1}^{q} (-1)^{k+1} s_k . 
\label{def_s_tilde}
\end{eqnarray}
      The action (\ref{action-NLSM}) is of the standard 
form of the NLSM: 
\begin{eqnarray}
       S_{\rm st} &=& \int^{\beta}_0 d\tau \int^{L}_0 dx 
\biggl\{ 
- i \frac{\theta }{4\pi} 
{\bf m} \cdot (\partial_{\tau} {\bf m} \times \partial_x {\bf m}) 
\nonumber \\ 
      &+& \frac{1}{2gv} (\partial_{\tau} {\bf m})^2 
+ \frac{v}{2g} (\partial_x {\bf m})^2 
\biggl\} , 
\label{standard_NLSM}
\end{eqnarray}
where $\theta$ is the topological angle, $g$ is the coupling constant 
and $v$ is the spin wave velocity. 

      The NLSM has a gapless excitation when $\theta/2\pi$ is 
a half-odd integer. 
      Comparing Eq.~(\ref{action-NLSM}) and Eq.~(\ref{standard_NLSM}), 
this condition is written in the gapless equation:  
\begin{equation}
\frac{1}{J^{(1)}} = \frac{2l-1}{4} \frac{1}{J^{(0)}} , 
\label{gapless_condition}
\end{equation}
where $l$ is an arbitrary integer. 
       For each $l$, this equation determines a 
boundary between disordered phases if it has a solution. 
       Note that Eq.~(\ref{gapless_condition}) is linear 
in \{$1/J_j$\}.

\section{Gapless lines for the spin chain with period 4}

       In the present case of Eqs. (\ref{magnitude}) and (\ref{exchange}), 
the transformation (\ref{transform}) is explicitly written as 
\begin{eqnarray}
{\bf n}_{4p+1} &=& \frac{3}{4}{\bf m}(p) 
+ \frac{1}{4}{\bf m}(p-1) + a {\bf L}_1(p) , 
\label{p_1}
\\
{\bf n}_{4p+2} &=& {\bf m}(p) + a {\bf L}_2(p) , 
\label{p_2}
\\
{\bf n}_{4p+3} &=& \frac{3}{4}{\bf m}(p) 
+ \frac{1}{4}{\bf m}(p+1) + a {\bf L}_3(p) , 
\label{p_3}
\\
{\bf n}_{4p+4} &=& \frac{1}{2}{\bf m}(p) 
+ \frac{1}{2}{\bf m}(p+1) + a {\bf L}_4(p) 
\label{p_4}
\end{eqnarray}
for the $p$th block. 
      Note that ${\bf m}(p-1)$ and ${\bf m}(p+1)$ are variables 
belonging to the neighboring blocks. 
      Since Eq.~(\ref{def_s_tilde}) is reduced to 
\begin{equation}
{\tilde s}_1 = s_a , \quad {\tilde s}_3 = s_b , 
\quad {\tilde s}_2 = {\tilde s}_4 = 0 , 
\label{s_tilde}
\end{equation}
then Eq.~(\ref{def_J_n}) yields 
\begin{eqnarray}
\frac{1}{J^{(0)}} &=& \frac{1}{4} \biggl( \frac{1}{J_{aa} s_a^2} 
+ \frac{2}{J_{ab} s_a s_b} + \frac{1}{J_{bb} s_b^2} \biggl) , 
\label{def_J_0} 
\\
\frac{1}{J^{(1)}} &=& \frac{1}{4} \biggl( \frac{1}{J_{aa} s_a} 
+ \frac{1}{J_{bb} s_b} \biggl) , 
\label{def_J_1}
\\
\frac{1}{J^{(2)}} &=& \frac{1}{4} \biggl( \frac{1}{J_{aa}} 
+ \frac{1}{J_{bb}} \biggl) . 
\label{def_J_2}
\end{eqnarray}
      Thus the NLSM action in the present case is given by Eq. 
(\ref{action-NLSM}) with Eqs. (\ref{def_J_0}) to (\ref{def_J_2}). 

      Substituting (\ref{def_J_0}) and (\ref{def_J_1}), 
the gapless equation (\ref{gapless_condition}) becomes 
\begin{equation}
\frac{1}{J_{ab}} = \frac{1 - t_a(l)}{2 t_b(l)} \frac{1}{J_{aa}} 
+ \frac{1 - t_b(l)}{2 t_b(l)} \frac{s_a}{s_b} 
\label{gapless_equation}
\end{equation}
with 
\begin{equation}
t_a(l) = \frac{2l-1}{4 s_a} , \quad t_b(l) = \frac{2l-1}{4 s_b} 
\label{ta-tb}
\end{equation}
in the unit of $J_{bb}=1$. 
       The gapless equation (\ref{gapless_equation}) represents a 
straight {\it gapless line} for each $l$ in the plane of 
$(1/J_{aa}, 1/J_{ab})$. 
       The part of a gapless line in the first quadrant is physical. 
       For negative $l$, Eq.~(\ref{gapless_equation}) does not produce 
a line which passes through the first quadrant. 
       Hence we hereafter consider $l$ as a positive integer. 

      When $s_a$ and $s_b$ are fixed, all the gapless lines with 
different values of $l$ intersect at a single point 
\begin{equation}
\biggl(  - \frac{s_a}{s_b} , 
\frac{1}{2} \biggl( 1 - \frac{s_a}{s_b} \biggl) \biggl) . 
\label{focus}
\end{equation}
      We call it the {\it focus};  this is in the second quadrant. 
      The slope of Eq.~(\ref{gapless_equation}) monotonically
decreases  as $l$ increases. 
       The maximum slope is $2s_b(1-1/4s_a)$ for $l=1$. 
       For large $l$, gapless lines accumulate toward the line 
\begin{equation}
\frac{1}{J_{ab}} = - \frac{s_b}{2s_a} \frac{1}{J_{aa}} 
- \frac{s_a}{2s_b} , 
\label{accumulate}
\end{equation}
which is outside the first quadrant. 
      The gapless lines for $l$ = 1, 2, $\cdots$, $2s_b$ are 
physical, because they go through the first quadrant. 
      The slope of a gapless line is positive if 1 $\le$ $l$ $\le$ 
$2s_a$,  and negative if $2s_a +1$ $\le$ $l$ $\le$ $2 s_b$. 
      A gapless line with smaller $l$ has a larger slope.

\section{Homogeneous spin magnitude} 

\begin{figure}[btp]
\begin{center}\leavevmode
\epsfxsize=90mm
\epsfbox{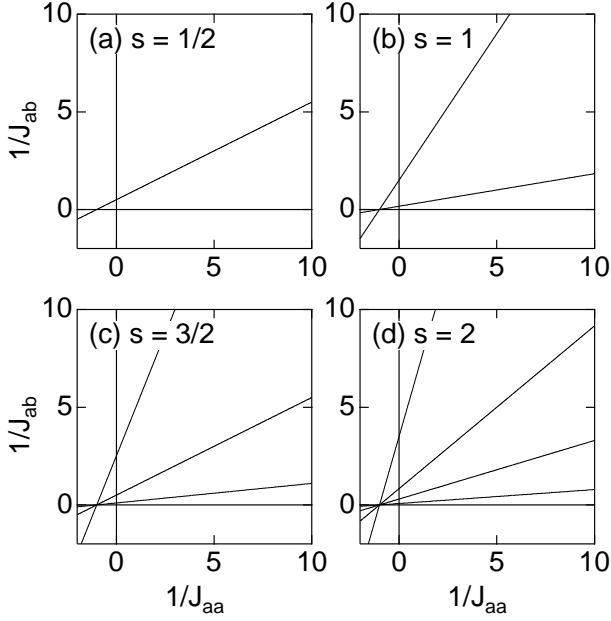}
\end{center}
\caption{ 
      Gapless lines for (a) $s$ = $\frac{1}{2}$, (b) 1, (c) $\frac{3}{2}$ 
and (d) 2 in the case of $s_a$ = $s_b$ ($\equiv$ $s$). 
      The unit of $J_{bb}=1$ is used. 
      The first quadrants represent the phase diagrams 
in the $1/J_{aa}$-$1/J_{ab}$ plane. 
      Phase boundaries are the gapless lines. 
} 
\label{gapless_homo}
\end{figure}

\begin{figure}[btp]
\begin{center}\leavevmode
\epsfxsize=90mm
\epsfbox{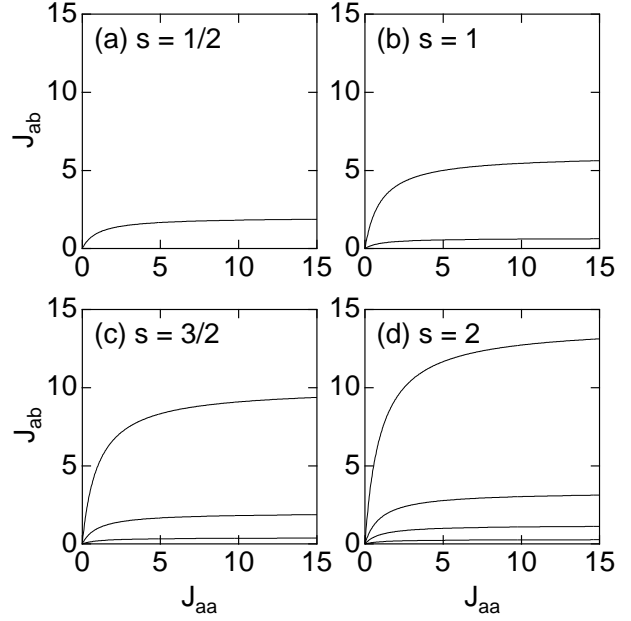}
\caption{ Phase diagrams in the $J_{aa}$-$J_{ab}$ plane 
for $s_a$ = $s_b$ ($\equiv$ $s$).  
The unit of $J_{bb}=1$ is used. }
\label{phase_homo}
\end{center}
\end{figure}

      In the case of $s_a = s_b \equiv s$, the spin magnitude is 
homogeneous and only the exchange coupling has a spatial 
modulation with period 4. 
      Then the gapless equation (\ref{gapless_equation}) is 
reduced to 
\begin{equation}
\frac{1}{J_{ab}} = - \frac{1}{2} \biggl( 1-\frac{4s}{2l-1} \biggl)
\biggl( \frac{1}{J_{aa}} + 1 \biggl) . 
\label{boundary_line_sa=sb}
\end{equation}
      This equation gives gapless lines for $l$ = 1, 2, $\cdots$ , 
$2s$ for each $s$. 
      Figure \ref{gapless_homo} shows them for 
$s$ = $\frac{1}{2}$, 1, $\frac{3}{2}$ and 2 in the 
$1/J_{aa}$-$1/J_{ab}$ plane. 
      The focus is $(-1, 0)$ for all $s$ and only the first quadrant is 
physical. 
      These gapless lines determine the phase boundaries among 
disordered phases. 
      The corresponding phase diagrams in the $J_{aa}$-$J_{ab}$ 
plane are shown in Fig.~\ref{phase_homo}. 
      When we consider the uniform case 
($J_{aa}$ = $J_{ab}$ =1 (= $J_{bb}$)), the gapless condition of 
Eq.~(\ref{boundary_line_sa=sb}) becomes $s = (2l-1)/2$, 
the Haldane's original result \cite{Haldane}.

      In the simplest case of $s=\frac{1}{2}$, 
Eq.~(\ref{boundary_line_sa=sb}) has a physical solution 
($J_{aa}>0$, $J_{ab}>0$) 
only for $l=1$ as shown in Fig.~\ref{gapless_homo}(a) and 
Fig.~\ref{phase_homo}(a). 
      Chen and Hida performed numerical calculation in this case 
and obtained a phase boundary \cite{Chen1}. 
      The gapless line given by the present NLSM method 
qualitatively agrees with their numerical phase boundary in the 
positive $J_{ab}$ region \cite{Chen2}. 

      The phases in Fig.~\ref{gapless_homo} or \ref{phase_homo} 
are interpreted by extending the valence-bond-solid (VBS) picture. 
      In the original VBS picture \cite{Affleck4}, a state with spin 
gap is represented by the direct product of singlet dimers. 
      Chen and Hida \cite{Chen1} extended the VBS picture to include 
local singlet states each of which is formed by four 
$\frac{1}{2}$-spins. 
      Using the extended VBS picture they explained their 
numerical phases in the $s$ = $\frac{1}{2}$ case. 
      We call the local singlet state the {\it singlet cluster} or 
simply the cluster. 
      We further extend the concept of the cluster to include a 
local singlet formed by more than four $\frac{1}{2}$-spins; 
      we will encounter larger clusters in the next section. 
      We refer the direct product state of regularly arrayed  
clusters as the {\it singlet cluster  solid} (SCS). 
      We can explain the disordered phases for arbitrary $s$ 
in the SCS picture. 

\begin{figure}[btp]
\begin{center}\leavevmode
\epsfxsize=65mm
\epsfbox{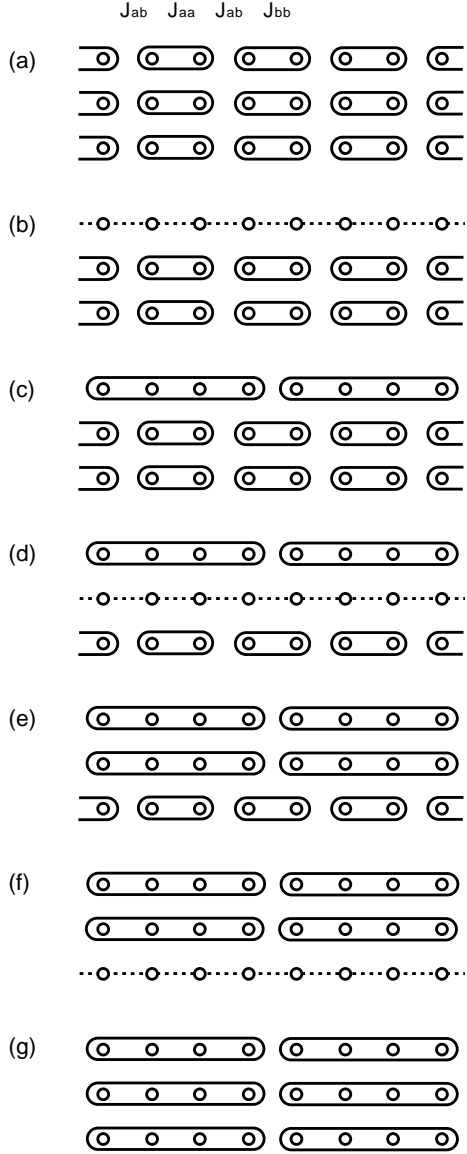}
\end{center}
\caption{The SCS picture for the phases in the case of 
$s$ = 3/2 (= $s_a$ = $s_b$). 
      The SCS states are labeled by (a) $C(6)$, (c) $C(4, 1)$, 
(e) $C(2, 2)$ and (g) $C(0, 3)$.}
\label{SCS_homo}
\end{figure}

      For definiteness, we explain the case of $s=\frac{3}{2}$ 
in the SCS picture;  the explanation for other values of $s$ 
is almost the same. 
      In Fig.~\ref{SCS_homo} the $s$=$\frac{3}{2}$ spin at each 
site is decomposed into three $\frac{1}{2}$-spins, 
which are represented by small circles. 
      The symmetrization of the wave function for the 
$\frac{1}{2}$-spins at a site retrieves the original 
$\frac{3}{2}$-spin. 
      A loop represents a local singlet state of two or four 
$\frac{1}{2}$-spins in it, while a dotted line represents a spatially 
extended singlet state. 
      Figures \ref{SCS_homo}(a), (c), (e) and (g) illustrate clockwise 
the phases in the first quadrant of Fig.~\ref{gapless_homo}(c). 
      We label them by $C(6)$, $C(4, 1)$, $C(2, 2)$ and $C(0, 3)$, 
respectively. 
      For example, $C(6)$ means that there are 6 singlet dimers 
in a unit cell, while $C(4, 1)$ means that there are 4 singlet dimers 
and 1 singlet cluster formed by four $\frac{1}{2}$-spins. 
      Figures \ref{SCS_homo}(b), (d) and (f) illustrate the states 
on the phase boundaries determined as the gapless lines with 
$l$ = 1, 2 and 3, respectively. 

      We first examine the phase $C(6)$ above the gapless line 
of $l$=1 in Fig.~\ref{gapless_homo}(c), which corresponds 
to the narrow region just above the $J_{aa}$-axis in 
Fig.~\ref{phase_homo}(c). 
      In this phase, $J_{ab}$ is much smaller than both $J_{aa}$ and 
$J_{bb}$ (=1). 
      Hence the ground state in the phase is the VBS state as shown 
in Fig.~\ref{SCS_homo}(a),  where singlet dimers are formed 
on $J_{aa}$- and $J_{bb}$-couplings. 
      As $J_{ab}$ increases, energy reduction by forming the dimers 
on the $J_{aa}$- and $J_{bb}$-couplings relatively decreases, 
since dimers on $J_{ab}$-couplings could reduce the energy. 
      When $J_{ab}$ arrives at a critical value, two singlet dimers 
per unit cell are broken and an extended state appears as shown in 
Fig.~\ref{SCS_homo}(b). 
      The extended singlet state explains the gapless line with 
the largest slope ($l$=1) in Fig.~\ref{gapless_homo}(c). 

\begin{figure}[btp]
\begin{center}\leavevmode
\epsfxsize=40mm
\epsfbox{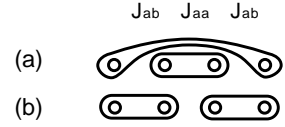}
\end{center}
\caption{Basis states (a) $| \ aa, bb >$ and (b) $| \ ba, ab >$ for 
a cluster of four $\frac{1}{2}$-spins (Eq.~(\ref{cluster4})). }
\label{cluster_basis}
\end{figure}

      When $J_{ab}$ increases further, the $\frac{1}{2}$-spins 
forming the spatially extended singlet state recombine to form 
singlet clusters of four $\frac{1}{2}$-spins. 
      Then the state becomes $C(4, 1)$ as shown in 
Fig.~\ref{SCS_homo}(c). 
      The formation of the cluster partially gain the exchange 
energy on the $J_{ab}$-couplings as well as that on the 
$J_{aa}$-coupling. 
      A singlet cluster of four $\frac{1}{2}$-spins is a linear 
combination of two basis states: 
\begin{equation}
\label{cluster4}
      | \ c^{(4)} > = w_1 | \ aa, bb > + w_2 | \ ba, ab > . 
\end{equation}
      Here $| \ aa, bb >$ is the direct product of two singlet 
pairs in the cluster;  
      one is a dimer pair of the adjacent $s_a$'s and the other is 
an extended pair of $s_b$'s as shown in Fig.~\ref{cluster_basis}(a). 
      The state $| \ ba, ab >$ is the direct product of two singlet 
dimers as shown in Fig.~\ref{cluster_basis}(b). 
      The relative weight $w_1/w_2$ of the basis states changes 
continuously when the exchange parameters changes within this 
phase region. 
      In contrast, the change from the state in Fig.~\ref{SCS_homo}(a) 
to that in Fig.~\ref{SCS_homo}(c) has occurred along with the 
disappearance of a series of singlet dimers and the appearance of 
a series of singlet clusters. 
       Hence the system has experienced a phase transition between 
these two states.  
      The critical state on the phase boundary is the gapless state 
in Fig.~\ref{SCS_homo}(b). 

      When $J_{ab}$ increases further from the state of 
Fig.~\ref{SCS_homo}(c), we can repeat the same explanation. 
      That is, a series of remnant singlet dimers are broken and 
singlet clusters increase as in Fig.~\ref{SCS_homo}(e) and 
finally as in Fig.~\ref{SCS_homo}(g). 
      The gapless extended states in Figs.~\ref{SCS_homo}(d) and 
\ref{SCS_homo}(f) appear as critical states in the way of changing. 
      They correspond to phase boundaries in 
Fig.~\ref{gapless_homo}(c) and Fig.~\ref{phase_homo}(c).

\section{Inhomogeneous spin magnitude}

      We examine the phase diagram for $s_a$$<$$s_b$. 
      The phase boundaries, i.~e. gapless lines, are given by 
Eq.~(\ref{gapless_equation}) for $l$ = 1, 2, $\cdots$, 2$s_b$. 

      The simplest example is the model with 
$s_a$ = $\frac{1}{2}$ and $s_b$ = 1. 
      In a previous paper, we examined this case and 
presented the phase diagram in the NLSM method \cite{Takano3}. 
      On the other hand, Tonegawa et al. \cite{Tonegawa1} and 
Hikihara et al. \cite{Tonegawa2} performed numerical calculation 
for the same case and obtained the numerical phase diagram. 
      The overall feature of our phase diagram well 
agrees with that of the numerical one in this example. 

\begin{figure}[btp]
\begin{center}\leavevmode
\epsfxsize=80mm
\epsfbox{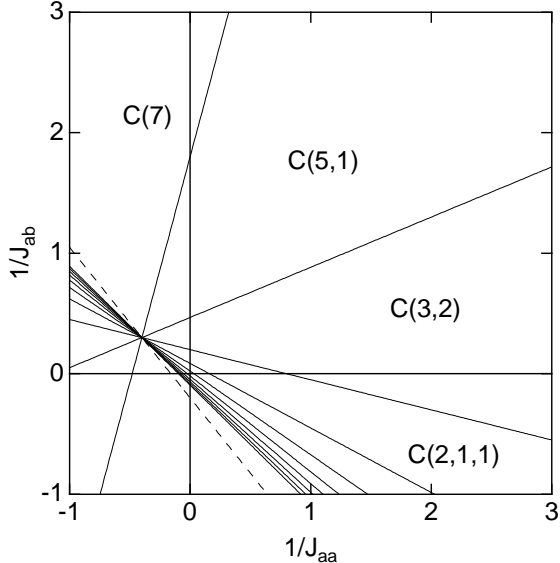}
\caption{ Gapless lines for $s_a = 1$ and $s_b = 5/2$. 
The lines with $l \le 10$ in Eq.~(\ref{gapless_equation}) are 
displayed, but only the lines with $l \le 5$ pass through the 
first quadrant. 
The dotted line is the large-$l$ limit of 
Eq.~(\ref{gapless_equation}). 
The first quadrant is the phase diagram in the 
$1/J_{aa}$-$1/J_{ab}$ plane. } 
\label{gapless_10_25}
\end{center}
\end{figure}

\begin{figure}[btp]
\begin{center}\leavevmode
\epsfxsize=80mm
\epsfbox{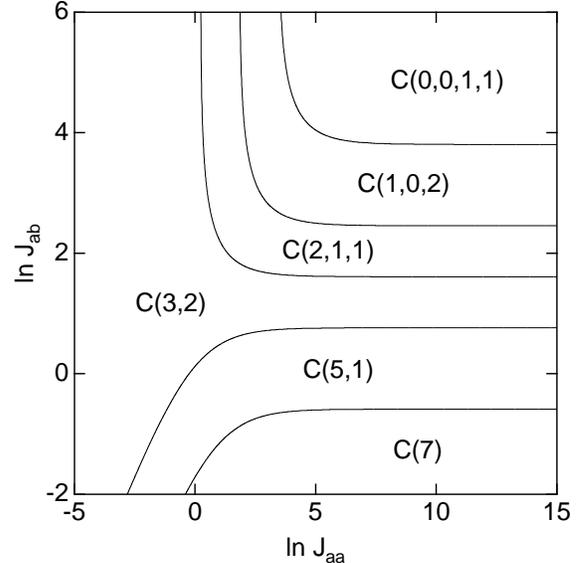}
\caption{Phase diagram for $s_a = 1$ and $s_b = 5/2$ 
in the $J_{aa}$-$J_{ab}$ plane. 
Logarithm is taken for each axis. }
\label{phase_10_25}
\end{center}
\end{figure}

      As a typical example with general features, 
we explain the case of 
$s_a$=1 and $s_b$=$\frac{5}{2}$ in detail. 
      The phase boundaries are given by Eq.~(\ref{gapless_equation}) 
with $l$ = 1, 2, $\cdots$, 5. 
      The phase diagram in the $1/J_{aa}$-$1/J_{ab}$ plane 
is shown in Fig.~\ref{gapless_10_25}, where 
only the first quadrant is physical. 
      All the gapless lines intersect on the focus 
($-\frac{2}{5}$, $\frac{3}{10}$) from Eq.~(\ref{focus}) 
and accumulate to the line 
$J_{ab}^{-1}$=$-\frac{5}{4}J_{aa}^{-1}$ $-\frac{1}{5}$ 
from Eq.~(\ref{accumulate}). 
      There are 6 phases in the first quadrant. 
      We label them clockwise as $C(7)$, $C(5, 1)$, $C(3, 2)$, 
$C(2, 1, 1)$, $C(1, 0, 2)$ and $C(0, 0, 1, 1)$;  the last two are not 
indicated in Fig.~\ref{gapless_10_25} because of the lack of space. 
      We will soon explain the general notation after introducing 
the SCS picture for the present example. 
      The corresponding phase diagram in the $J_{aa}$-$J_{ab}$ plane 
is shown in Fig.~\ref{phase_10_25}. 
      We have taken the logarithms for $J_{aa}$ and $J_{ab}$ 
because of the largely different scales of the phases. 

\begin{figure*}
\begin{center}\leavevmode
\epsfxsize=150mm
\epsfbox{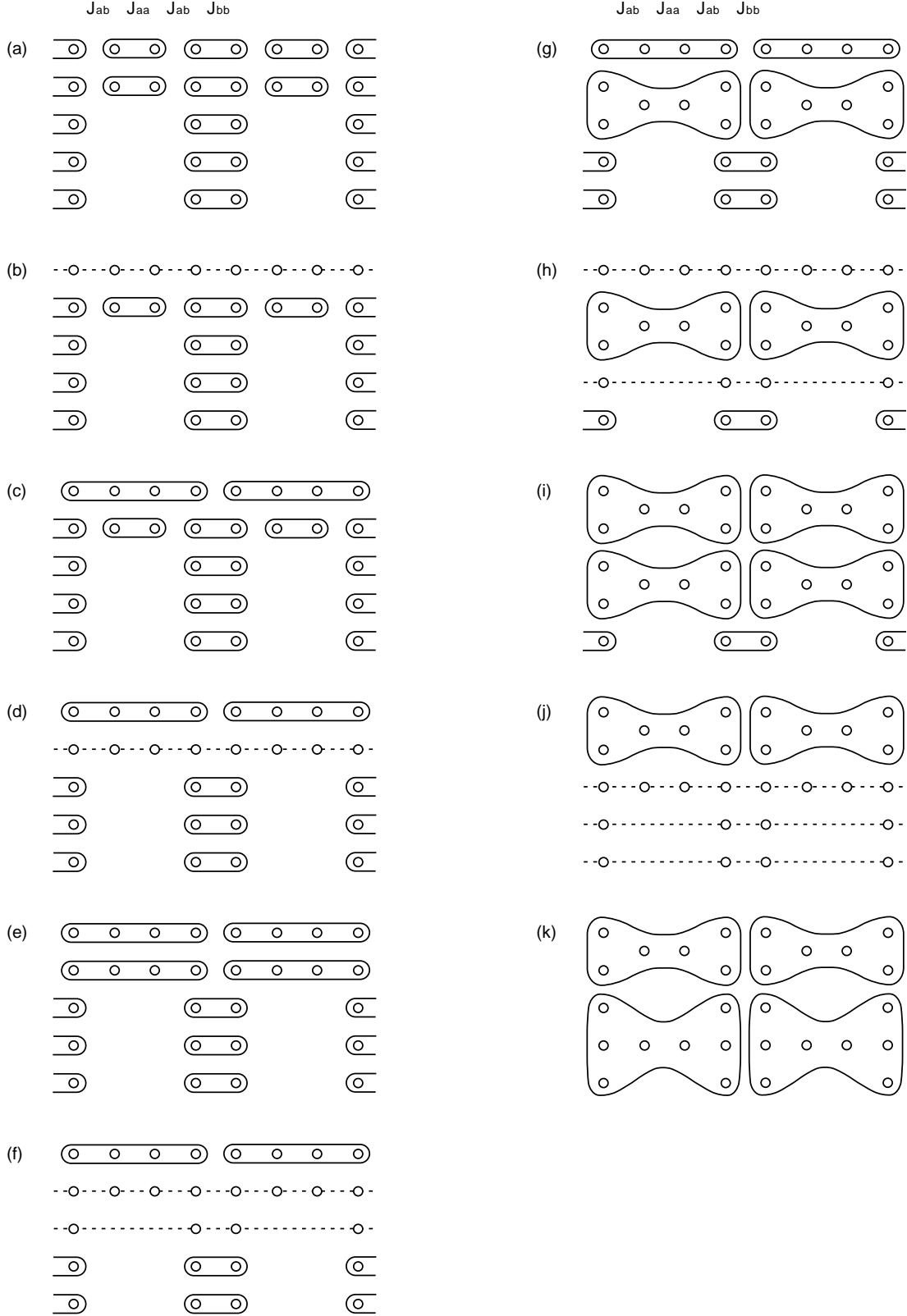}
\end{center}
\caption{The SCS picture for $s_a$=1 and $s_b$=$\frac{5}{2}$. 
      Gapful phases in Fig.~\ref{gapless_10_25} or 
\ref{phase_10_25} are represented by 
(a) $C(7)$, (c) $C(5, 1)$, (e) $C(3, 2)$, (g) $C(2, 1, 1)$, (i) $C(1, 0, 2)$ 
and (k) $C(0, 0, 1, 1)$. 
      The other figures represent gapless states on phase 
boundaries.}
\label{SCS_hetero}
\end{figure*}

      We interpret the gapful phases in the SCS picture. 
      We first decompose the spin (with magnitude $s_a$ or $s_b$) 
at each site into $\frac{1}{2}$-spins in the same manner 
as in Sec.~V. 
      The original spin is retrieved by symmetrizing the wave 
function for the $\frac{1}{2}$-spins at the same site. 
      We represent a gapful phase as an SCS state, which is a 
direct product of singlet clusters. 
      The SCS states for $s_a$=1 and $s_b$=$\frac{5}{2}$ are 
illustrated in Fig.~\ref{SCS_hetero}. 
      Each decomposed $\frac{1}{2}$-spin is represented by a small 
circle. 
      A loop containing an even number of $\frac{1}{2}$-spins 
means a singlet cluster, 
while a dotted line means an extended singlet state. 
      The general SCS state is labeled as 
\begin{equation}
\label{cluster}
C(k_1, k_2, k_3, \cdots ) , 
\end{equation}
where $k_j$ is the number of singlet clusters formed by $2j$ 
$\frac{1}{2}$-spins in a unit cell. 
      SCS states of $C(k_1)$ and $C(k_1, k_2)$ have been introduced 
in Sec. V. 
      For example, a unit cell of $C(0, 0, 1, 1)$ includes a singlet 
cluster consisting of four $\frac{1}{2}$-spins and a singlet 
cluster consisting of six $\frac{1}{2}$-spins. 

\widetext

\narrowtext

      We explain the phases in the first quadrant of 
Fig.~\ref{gapless_10_25} clockwise starting from the phase 
$C(7)$. 
      In this phase, $J_{ab}$ is much smaller than both $J_{aa}$ and 
$J_{bb}$ (=1) so that a singlet dimer is not formed on any 
$J_{ab}$-coupling. 
      Hence the $\frac{1}{2}$-spins form singlet dimers on $J_{aa}$- 
and $J_{bb}$-couplings to gain exchange energies. 
      This ground state is explained as a VBS state illustrated in 
Fig.~\ref{SCS_hetero}(a). 
      When $J_{ab}$ increases, the system experiences the transitions 
from $C(7)$ to $C(5, 1)$ and from $C(5, 1)$ to $C(3, 2)$. 
      Then singlet dimers decrease and singlet clusters consisting of 
four $\frac{1}{2}$-spins increase in each unit cell. 
      The ground states of $C(5, 1)$ and $C(3, 2)$ are illustrated in, 
respectively, (c) and (e) of Fig.~\ref{SCS_hetero}. 
      The states in (b) and (d) are on the phase boundaries. 
      Extended singlet states appear just when a series of singlet 
dimers are broken and a series of singlet clusters are formed. 
      Gapless spin excitations originate from the extended states. 
      Until now the explanation is the same as that for the homogeneous 
case in Sec. V. 

      The region of $C(3, 2)$ in the $J_{aa}$-$J_{ab}$ plane 
(Fig.~\ref{phase_10_25}) is T-shaped;  the T is rotated by 90 
degrees. 
      We explain the origin of this shape by Fig.~\ref{SCS_hetero}(e). 
      A singlet cluster state of four $\frac{1}{2}$-spins in 
this figure is described by Eq. (\ref{cluster4}). 
      First, in the large $J_{aa}$ part of this T-shaped region, 
$J_{aa}$ is much larger than the others. 
      Hence, in a cluster, the state $| \ aa, bb >$ including a dimer 
on the $J_{aa}$-coupling is more favorable than $| \ ba, ab >$; i.~e. 
the weight $w_1$ is much larger than $w_2$ in Eq. (\ref{cluster4}). 
      Second, in the large $J_{ab}$ part of the T-shaped region, 
$J_{ab}$ is larger than $J_{aa}$ and $J_{bb}$ (=1). 
      Hence the state $| \ ba, ab >$ is favorable since it 
consists of two singlet dimers on $J_{ab}$-couplings; i.~e. 
the weight $w_2$ is much larger than $w_1$. 
      Third, in the small $J_{ab}$ part of the T-shaped region, 
$J_{bb}$ (=1) is larger than the others. 
      Because of the lack of a $J_{bb}$-coupling within a cluster, 
there is no particularly favorable dimers in Eq. (\ref{cluster4}). 
      The state $| \ aa, bb >$ or $| \ ba, ab >$ individually gains 
the exchange energy to some extent. 
      However comparable values of $w_1$ and $w_2$ further reduce 
the total energy because of the off-diagonal matrix element 
$< aa, bb \ | H | \ ba, ab >$. 
      Thus the above three parts are differently characterized, 
but they belong to the same phase and merge 
around ($J_{aa}$, $J_{ab}$) = (1, 1). 
      One can move from one part to another 
by continuously changing the weights, $w_1$ and $w_2$, 
with no phase transition. 

      When $J_{aa}$ and $J_{ab}$ become larger, the system 
further intends to gain the exchange energies on $J_{aa}$- 
and $J_{ab}$-couplings rather than on $J_{bb}$-couplings. 
      Then the ground state $C(3, 2)$ changes into $C(2, 1, 1)$. 
      In $C(2, 1, 1)$, singlet clusters consisting of six 
$\frac{1}{2}$-spins appear as shown in Fig.~\ref{SCS_hetero}(g). 
      At the moment of the transition from $C(3, 2)$ to $C(2, 1, 1)$, 
a series of singlet dimers on $J_{bb}$-couplings break and an 
extended state appears as shown in Fig.~\ref{SCS_hetero}(f). 
      The gapless phase boundary in Fig.~\ref{gapless_10_25} and 
Fig.~\ref{phase_10_25} results from the extended states. 
      As $J_{aa}$ and $J_{ab}$ become further large, singlet dimers 
 on $J_{bb}$-couplings become much more unfavorable for 
the energy reduction. 
      Instead the system reduces the total energy by forming more or 
larger singlet clusters. 
      Hence the ground states in Fig.~\ref{SCS_hetero}(i) and (k) 
appear as both $J_{aa}$ and $J_{ab}$ increase. 
      These ground states correspond to the phases $C(1, 0, 2)$ and 
$C(0, 0, 1, 1)$. 
      There appear gapless extended states (h) and (j) 
of Fig.~\ref{SCS_hetero} at the transitions 
from (g) to (i) and from (i) to (k), respectively. 
      They correspond to the phase boundaries between 
$C(2, 1, 1)$ and $C(1, 0, 2)$, and between $C(1, 0, 2)$ and 
$C(0, 0, 1, 1)$. 

      In the general case of arbitrary $s_a$ and $s_b$ 
($s_a$$<$$s_b$), the number of phases may be large but 
the structure of the phase diagram is essentially the same. 
      There are 2$s_b$+1 phases divided by 2$s_b$ phase boundaries 
or gapless lines, which are given by Eq.~(\ref{gapless_equation}) 
for $l$ = 1, 2, $\cdots$, 2$s_b$. 
      The phase below the line of $l$=1 in the $J_{aa}$-$J_{ab}$ 
plane is interpreted as a VBS state $C(2s_b+2s_a)$ consisting only 
of singlet dimers like Fig.~\ref{SCS_hetero}(a). 
      As $J_{ab}$ increases, the system successively undergoes 
the phases $C(2s_b+2s_a-2, 1)$, $C(2s_b+2s_a-4, 2)$, $\cdots$ , 
$C(2s_b-2s_a, 2s_a)$. 
      For each phase change, two singlet dimers disappear and 
a cluster with four $\frac{1}{2}$-spins appears in each unit cell. 
      The ($2s_a +1$)th phase $C(2s_b-2s_a, 2s_a)$ is T-shaped 
like $C(3, 2)$ of Fig.~\ref{phase_10_25}. 
      As $J_{ab}$ further increases with $J_{aa}$, the number of 
the singlet dimers continues to decrease one by one; for a single 
phase change a singlet dimer disappears and a larger cluster 
with more than four $\frac{1}{2}$-spins is formed in each unit 
cell. 
      Hence there are $2s_b-2s_a$ phases above the boundary 
line of $l=2s_a+1$ in the $J_{aa}$-$J_{ab}$ plane. 
      In the phase of the large limit of $J_{ab}$, there are $2s_a$ 
clusters like Fig.~\ref{SCS_hetero}(k). 
      Hence a cluster in this phase includes $2(s_a+s_b)/s_a$ 
$\frac{1}{2}$-spins in average. 
      That is, there are clusters with $[2s_b/s_a]+2$ 
$\frac{1}{2}$-spins and clusters with $[2s_b/s_a]+3$ 
$\frac{1}{2}$-spins; 
      here $[\cdots]$ means the integer part of the number inside. 
      Thus we have explained the 2$s_b$ phase transitions which 
appear with increasing $J_{ab}$ (and $J_{aa}$).

\section{Summary and Discussion}

      We applied the previously developed NLSM method to the mixed 
spin chain consisting of spins with magnitudes $s_a$ and $s_b$ 
($s_a \le s_b$). 
      They are arrayed as $s_a$-$s_a$-$s_b$-$s_b$ and the exchange 
parameters are periodic with period 4. 
      The NLSM for this spin chain yields the gapless equation 
which determines the 2$s_b$ phase boundaries. 
      Thus we obtained the phase diagram in the 
exchange parameter space. 
      The number of the phases is 2$s_b$+1 irrespective of $s_a$. 
      To explain the ground states in the phase diagram 
we proposed the SCS picture, 
which is an extension of the VBS picture. 
      In the SCS picture, the gapful ground state in each phase is 
represented as a direct product of regularly arrayed singlet 
clusters. 
      Then the phase transition is a change from an SCS state 
to another and a gapless state appears just at the change. 
      The SCS picture qualitatively explains, in a unified way, 
all the phases which were  obtained by the present NLSM method 
for arbitrary $s_a$ and $s_b$. 

      The phase diagrams for simple cases were compared to 
results of numerical calculations. 
      The features of the phase boundaries qualitatively agree with 
those of the numerical calculations for $s_a$=$s_b$=$\frac{1}{2}$ 
\cite{Takano1,Chen1} and for $s_a$=$\frac{1}{2}$ and $s_b$=1 
\cite{Takano3,Tonegawa1,Tonegawa2}. 
      It is expected that numerical calculations will be performed 
for various $s_a$ and $s_b$. 
      Experiments for real materials have not been seen yet. 
      It is also expected that materials with various $s_a$ and $s_b$ 
will be found and experiments will be performed for them.

\section*{Acknowledgment}

      I thank Kazuo Hida for useful discussion. 
      This work is supported by the Grant-in-Aid for 
Scientific Research from the Ministry of Education, Science, 
Sports and Culture, Japan. 



\begin{references}
\bibitem{even_block} 
      The block must be taken as an even number of 
unit cells if a unit cell consists of an odd number  of spins.
\bibitem{Chen2} 
      They parameterized the exchange parameters as 
$J_{aa} \equiv 1 - \delta$, $J_{bb} \equiv 1 + \delta$ and 
$J_{ab} \equiv j$. 
      Then the equation (\ref{boundary_line_sa=sb}) with 
$s=\frac{1}{2}$ and $l=1$ reduces to 
$\delta = (1 - j)^{1/2}$. 
      This equation and their numerical result are compared in a 
figure of their paper \cite{Chen1}.
\end{references}
\end{document}